\documentclass[aps,prb,preprint,superscriptaddress,showpacs]{revtex4-1}

\usepackage{textcomp}
\usepackage{amsmath}
\usepackage{amssymb}
\usepackage{bm}
\usepackage{color}
\usepackage{dcolumn} 
\usepackage[caption=false]{subfig} 
\usepackage{footnote}
\usepackage{enumerate}
\usepackage{epstopdf}

\usepackage[]{graphicx} 
\usepackage{bmpsize}

\usepackage{hyperref}
\newcommand{\kvec}{$\mathbf{k}$}
\newcommand{\qvec}{$\mathbf{q}$}

\newcolumntype{d}[1]{D{.}{.}{#1}}
\newcolumntype{e}[1]{D{.}{}{#1}}

\newcommand{\qe}{{\sc Quantum ESPRESSO}}
\newcommand{\qeabbrev}{{\sc QE}}

\begin{document}

\title{On the possibility of metastable metallic hydrogen}

\author{Craig M.\ Tenney}
\affiliation{Department of Physics and Astronomy, Washington State University, Pullman, Washington 99164, USA}

\author{Keeper L.\ Sharkey}
\affiliation{Department of Physics and Astronomy, Washington State University, Pullman, Washington 99164, USA}

\author{Jeffrey M.\ McMahon}
\email[]{jeffrey.mcmahon@wsu.edu}
\affiliation{Department of Physics and Astronomy, Washington State University, Pullman, Washington 99164, USA}

\date{\today}


\begin{abstract}
Metallic hydrogen is expected to exhibit remarkable physics. Examples include high-temperature superconductivity and possible novel types of quantum fluids. These could have revolutionary practical applications. The pressures required to obtain metallic hydrogen, however, are expected to be significant. For practical, and terrestrial applications, a key question is therefore whether this phase is metastable at lower pressures. In this work, this possibility is investigated, using first-principles simulations. The results show that metallic hydrogen is metastable, but strongly suggest that it is not so to ambient conditions. Implications of these results for fundamental physics, and also practical applications of metastable metallic hydrogen are discussed.
\end{abstract}





\maketitle



In 1935, Wigner and Huntington predicted \cite{:/content/aip/journal/jcp/3/12/10.1063/1.1749590} that sufficient pressure would dissociate hydrogen molecules; and any Bravais lattice of such atoms would be metallic. Initial interest in metallic hydrogen was primarily related to astrophysical problems \cite{0034-4885-73-1-016901}. Eventually, and to more recently though, there has been significant interest in this phase at relatively low temperatures \cite{RevModPhys.84.1607}. This can be attributed to the remarkable properties that are expected. This includes, for example, high-temperature superconductivity \cite{PhysRevLett.21.1748, PhysRevB.84.144515, PhysRevB.85.219902}. The possibility of a quantum liquid ground-state has also been suggested \cite{Bonev2004}; in which case, hydrogen may have quantum ordered states that represent novel types of quantum fluids \cite{Babaev2004}. Applications of the (expected) remarkable physics could revolutionize several fields; such as energy, technology, and rocketry \cite{1742-6596-215-1-012194}. 

The pressures involved to dissociate hydrogen molecules are expected to be significant though [$447(3)$ GPa \cite{PhysRevLett.114.105305}]. Particularly important for practical, and terrestrial applications is therefore whether metallic hydrogen is metastable. This possibility was suggested in 1972 \cite{Brovman1972, 0038-5670-14-6-A30}. However, it has received little \cite{:/content/aip/journal/jap/111/6/10.1063/1.3694793} to no attention since then; both at low (to zero) pressures and higher (but below molecular dissociation). 

The theoretical capabilities available in 1972, however, limited the calculations \cite{Brovman1972} to an approximate, perturbative treatment; and consideration of only one- (and few two-) atom lattices. Theoretical capabilities have significantly progressed \cite{RevModPhys.84.1607} though; properties can now be accurately calculated from first principles. Considering this problem with modern theoretical methods is therefore critical for addressing this gap in knowledge; and also providing important information for experiments. This is particularly timely, with recent claims \cite{Diaseaal1579} (see also Refs.\ \onlinecite{2017arXiv170204246G, 2017arXiv170205125E, 2017arXiv170207192L, 2017arXiv170303064S}) of the production of metallic hydrogen at low temperatures in static, diamond anvil cell experiments.

The answer to the question of metastability presupposes solution to a series of interrelated problems:
\begin{enumerate}[{(1)}]
    \item Determination of the minimum-energy crystal structure(s). This includes proof that they lie at stationary points (with respect to all parameters).
    \item Proof of (dynamic) stabilities.
    \item Analysis of the relation between the ground- and metastable-state structure(s) (assuming that such exist), and the molecular phase. This includes that of the processes that are undergone during transitions between them, and also when the pressure is removed.
    \item Determination of the lifetime of this phase.
\end{enumerate}

In this Article, we present results from first-principles simulations in support of this answer. Particular focus is given to problems ($1$) and ($2$) above; and indirectly to ($3$). A discussion of these results is given, and conclusions are drawn; including implications for fundamental physics, and also practical applications. A discussion of the methods used follows the main text. A Supplementary Information (SI) accompanies this work.

\section{Results}
\label{sec:results}

\subsection{Body-Centered Tetragonal Structures}

\begin{figure*}
    \includegraphics[width=0.5\textwidth]{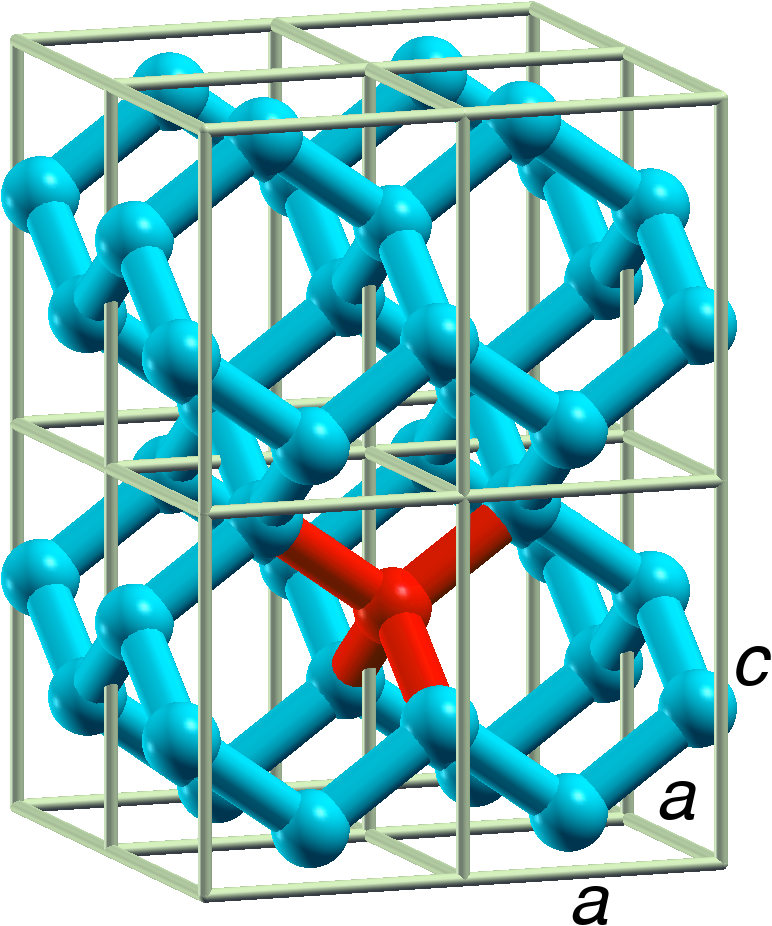}
    \caption{Body-centered tetragonal (BCT) structure (I$4_1/$amd space group) of atomic metallic hydrogen. Shown is a $2{\times}2{\times}2$ supercell. Notice the tetragonal coordination of atoms (lower, central atom highlighted). (Fictitious bonds have been drawn for clarity.) Notice also the square base ($a{\times}a$) and height ($c \neq a$).}
    \label{fig:tetragonal}
\end{figure*}

The most promising candidate structures for metastable metallic hydrogen (at least at higher pressures; and likely what one would find experimentally) are the most stable ones just above molecular dissociation. While these are not known experimentally, first-principles simulations \cite{PhysRevLett.106.165302} suggest that they can be represented by a body-centered tetragonal (BCT) structure (possibly including lower-symmetry distortions \cite{:/content/aip/journal/jap/111/6/10.1063/1.3694793}); as shown in Fig.\ \ref{fig:tetragonal}.

In this representation, there are five structures of note (in atomic hydrogen), defined by the axial ratio $c/a$; three of which have been considered before: those of $\beta$-Sn ($c/a < 1$), diamond ($c/a = \sqrt{2}$), and Cs-IV ($c/a > 1$), and the other two with $c/a \ll 1$ and $\gg 1$. Their structural parameters are reported in the SI. An analysis of (only) these structures reveals several general, qualitative features about metastable metallic hydrogen.

\begin{figure}
    \includegraphics[width=0.5\textwidth]{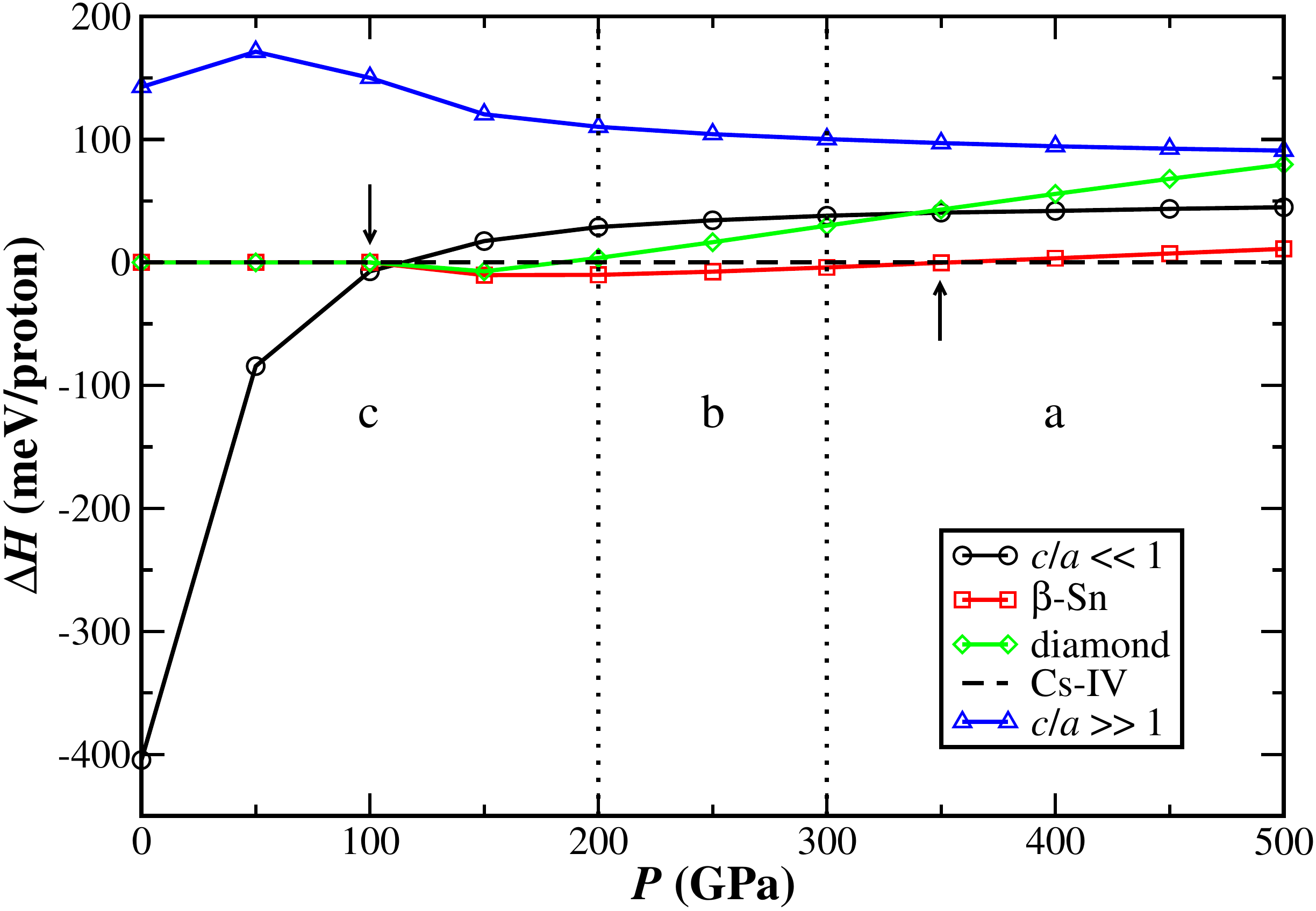}
    \caption{Phase diagram (at $0$ K) of the BCT structures, below molecular dissociation. Enthalpies $\Delta H$ are shown relative to Cs-IV, as a function of pressure $P$. Three (qualitative) regions of stability are separated by dotted lines, and labeled a, b, c. Up and down arrows are used to indicate where the $\beta$-Sn and $c/a \ll 1$ structures become the most stable, respectively. The latter arrow also points to where the $\beta$-Sn, diamond, and Cs-IV structures become degenerate in enthalpy.}
    \label{fig:enthalpies}
\end{figure}

Consider the phase diagram (at $0$ K), below molecular dissociation; shown in Fig.\ \ref{fig:enthalpies}.
The pressure range can be (qualitatively) divided into three regions:
\begin{enumerate}[{(a)}]
    \item $> 300$ GPa: the relative enthalpies of all of the structures remain relatively flat; the only exception being the diamond structure (discussed in more detail below).  
    \item $200$--$300$ GPa: the enthalpies of the $\beta$-Sn, diamond, and Cs-IV structures become nearly degenerate; while those with $c/a \ll 1$ and $\gg 1$ begin to decrease and increase, respectively.
    \item $< 200$ GPa: the $\beta$-Sn, diamond, and Cs-IV structures become degenerate (precisely at $100$ GPa); while those with $c/a \ll 1$ and $\gg 1$ sharply decrease and increase, respectively.
\end{enumerate}
This division (as meaningful) is supported by the results of additional calculations; presented and discussed below, and also in Section I\! B\! $1$ of the SI. 

\begin{figure}
    \includegraphics[width=0.5\textwidth]{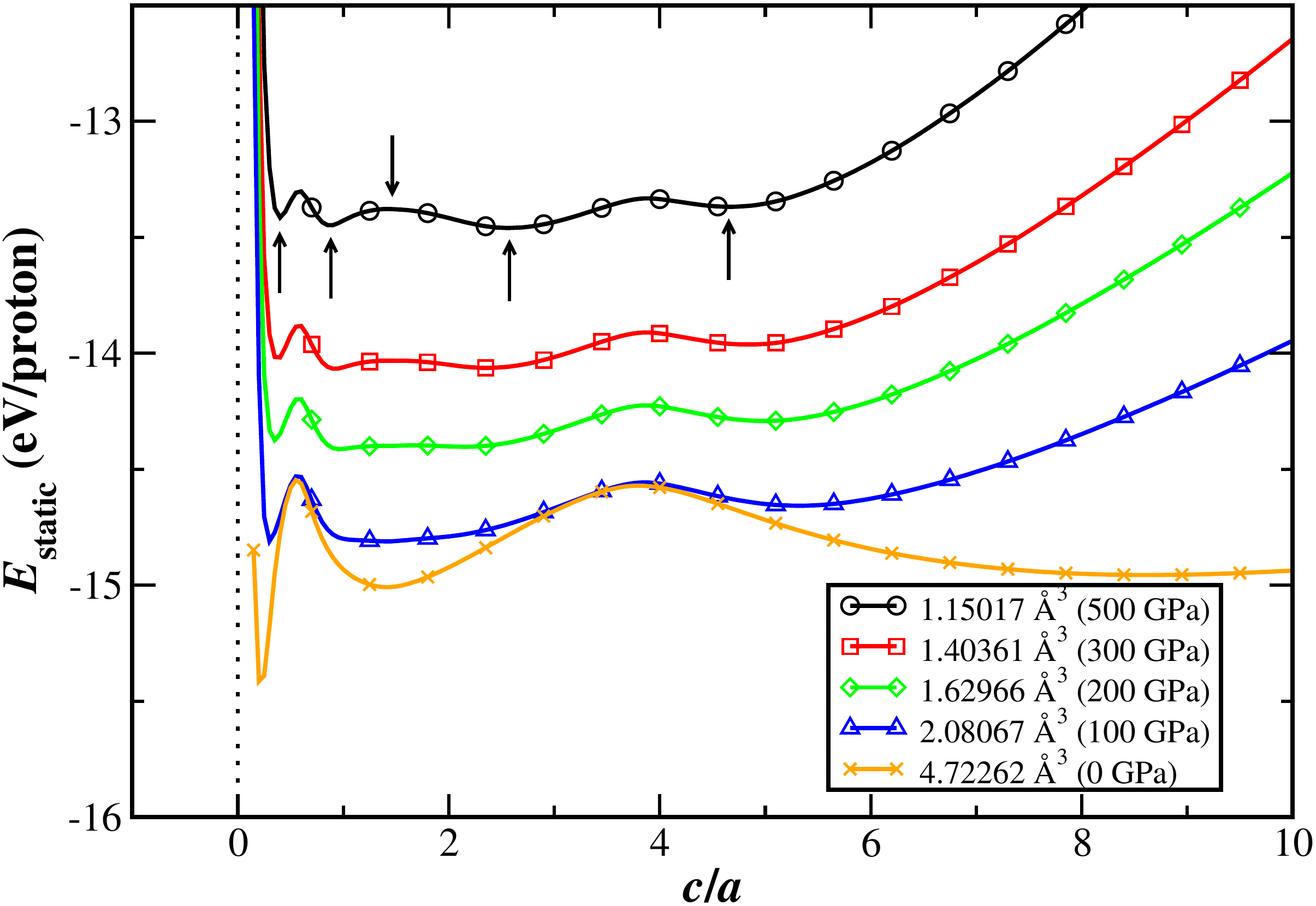}
    \caption{Energies (static) $E_\text{static}$ of the BCT structure, as a function of the axial ratio $c/a$. Fixed volumes are shown, calculated as an average of each structure at the pressure indicated in parenthesis. The $c/a \ll 1$, $\beta$-Sn, diamond, Cs-IV, and $\gg 1$ structures are indicated with arrows (in order of increasing $c/a$), at $500$ GPa.}
    \label{fig:c_a}
\end{figure}

Additional insight (including into Fig.\ \ref{fig:enthalpies}) is obtained by considering the energy of the BCT structure over the entire range of $c/a$; shown in Fig.\ \ref{fig:c_a}. 
As will be supported by the results of additional calculations (below), there are two especially important points that are suggested by these results:

First: \textit{the collapse of energy barriers in the potential-energy surface (PES) of (atomic) metallic hydrogen, between $300$ and $200$} GPa. Consider the diamond structure. At relatively high pressures, this structure forms an energy barrier (a maximum) between those of $\beta$-Sn and Cs-IV. While Fig.\ \ref{fig:enthalpies} shows that its relative enthalpy decreases continuously with decreasing pressure, it is between $300$ and $200$ GPa that it (as an energy barrier) does so appreciably. And, by $100$ GPa, it has collapsed completely, leaving only a single minimum in this region of $c/a$ ($\approx \sqrt{2}$). Notice in Fig.\ \ref{fig:enthalpies} that the $\beta$-Sn, diamond, and Cs-IV structures here become degenerate in enthalpy.

Second: \textit{the tendency towards the formation of molecules, below $200$} GPa. Consider the structures with $c/a \ll 1$ and $\gg 1$. The sharp increases in energy as $c/a \rightarrow 0$ and $\rightarrow \infty$ are due to the Coulomb repulsion resulting from Brillouin planes with like charges that approach each other. This (approach) though is also related to the formation of molecules; resulting in a minimum of energy, prior to its sharp increase. This explains the sharp decrease in energy of the $c/a \ll 1$ structure. Note the nearest-neighbor distance (at $0$ GPa) of $0.9900$ {\AA}.

The ability of atoms to approach each other is the result of the collapse of (atomic) energy barriers (discussed above). While this occurs, there is also a hardening of molecular ones. Notice the relative heights (and their trends with pressure) of the maxima separating the $c/a \ll 1$ and $\gg 1$ structures from the atomic ($c/a \approx \sqrt{2}$) region.

\subsection{Structure Prediction}

While the BCT structures (discussed above) are the most promising candidates at pressures near molecular dissociation, there is no guarantee that they are representative of those (potentially) so at lower ones. Consider the lattice energy $E_\text{lattice}$ of metallic hydrogen; in the adiabatic approximation:
\begin{equation}
    E_\text{lattice} = E_\text{static} + E_\text{vib} ~~~ ,
\end{equation}
where $E_\text{static}$ is the static (electronic ground-state energy, with a fixed lattice) contribution, and $E_\text{vib}$ is that resulting from atomic vibrations. In metallic hydrogen (and unlike ordinary metals), $E_\text{static}$ favors anisotropic structures; see Ref.\ \onlinecite{Brovman1972} for a discussion. $E_\text{vib}$, on the other hand, favors symmetric ones. The subtle (and important) details of these terms, including their relative magnitudes, depends on the pressure. At low pressures, $E_\text{lattice}$ is determined largely by $E_\text{static}$. It is with increasing pressure that $E_\text{vib}$ becomes (especially) important.

Searches were performed for metallic (atomic and/or mixed atomic/molecular) structure(s) with the lowest $E_\text{static}$ at $0$ GPa. These revealed several candidate structures of (potentially) metastable metallic hydrogen. An analysis of the most stable ones over the entire pressure range further supports the results above; such is provided in Section I\! B\! $1$ of the SI. At present, however, it is relatively low pressures that are of main interest.

Below $200$ GPa, \textit{all} structures found show a tendency towards the formation of molecules. This is consistent with the (second) point suggested by Fig.\ \ref{fig:enthalpies} (above). It is supported though by a detailed analysis (below) of the structures themselves.

In searches with more than a few atoms, \textit{only} molecular or mixed atomic/molecular structures were found. In addition, for the latter, as the number of atoms increases, there is a greater proportion of molecules (to atoms). This trend is also reflected in their relative enthalpies. Results and discussion for the most stable of these structures are provided in Section I\! B of the SI. That this trend occurs means that such structures may not be at \textit{stable} stationary points.

\begin{figure}
    \centering
    \subfloat[Pmmm]{{\includegraphics[width=3.075cm]{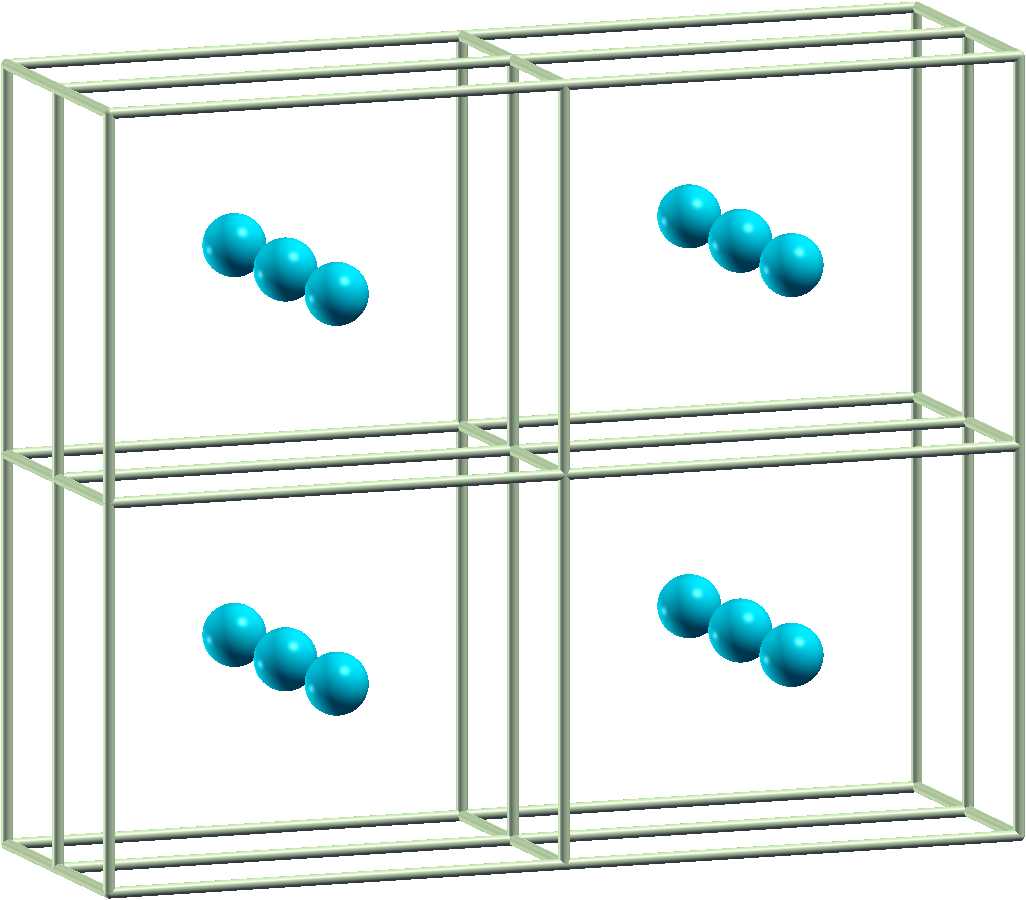}}}
    \qquad
    \subfloat[Cmmm]{{\includegraphics[width=3.5cm]{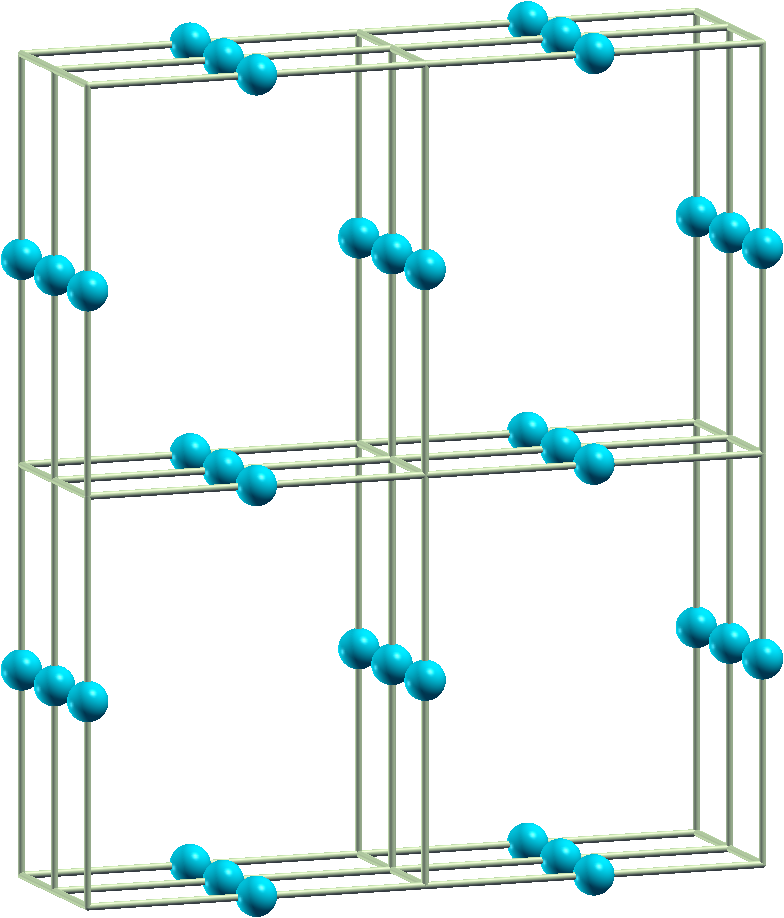}}}
    \qquad
    \subfloat[Immm]{{\includegraphics[width=2.85cm]{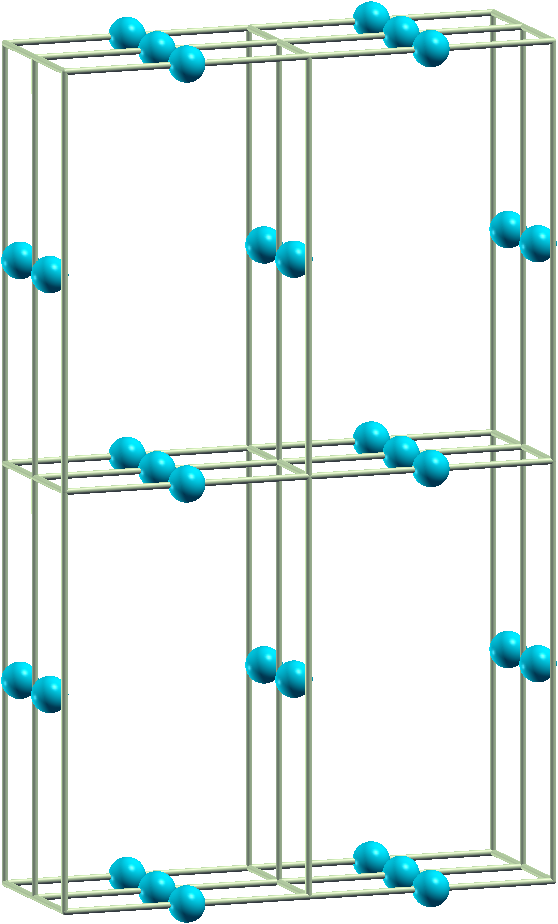}}}
    \qquad
    \subfloat[P$6/$mmm]{{\includegraphics[width=4.05cm]{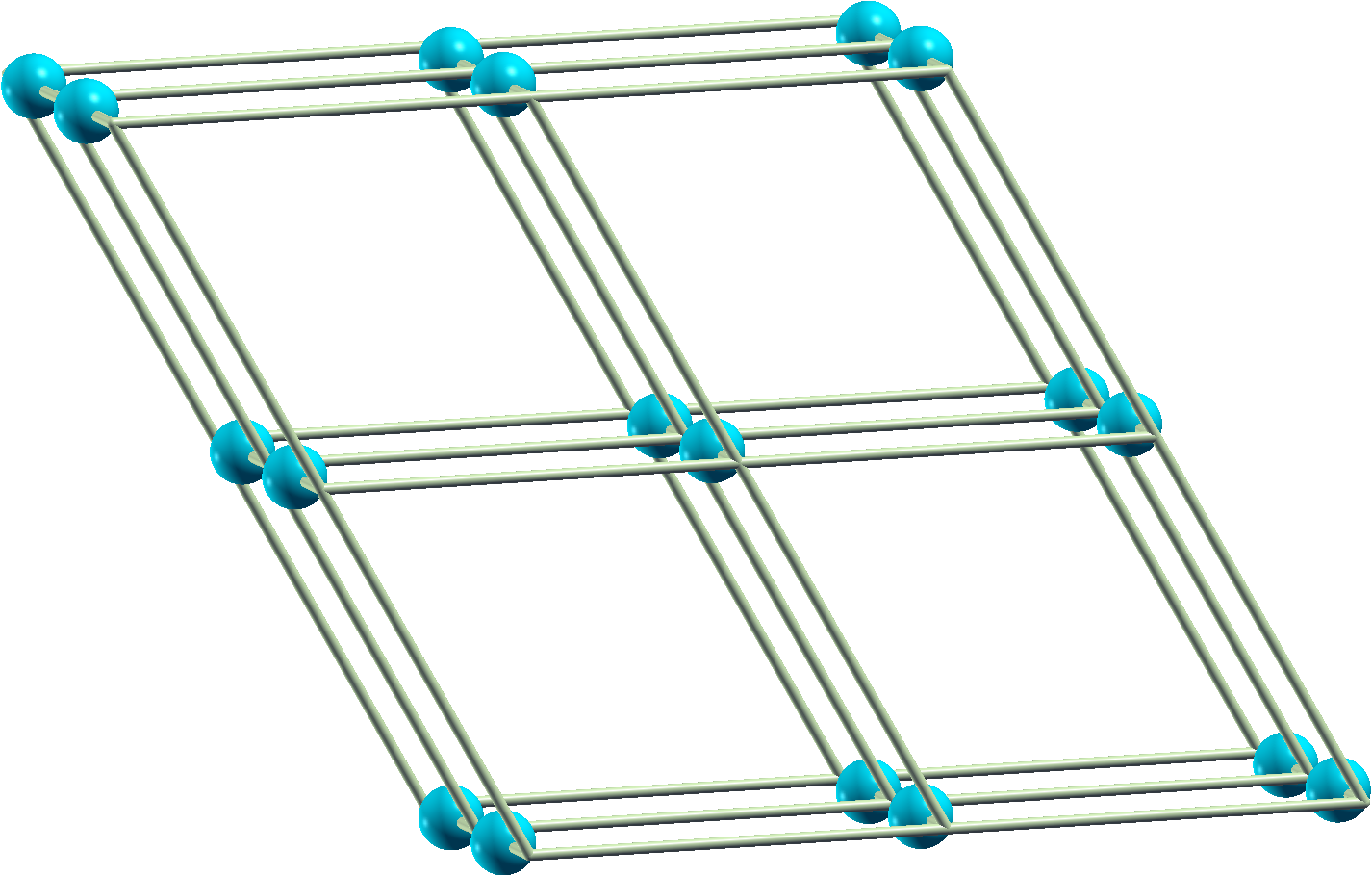}}}        
    \caption{Most stable candidate structures of metastable (atomic) metallic hydrogen, at $0$ GPa. Shown are $2{\times}2{\times}2$ supercells. Their space groups are specified in the subcaptions.}
    \label{fig:atomic_structures}
\end{figure}

The results above suggest that any metastable structure(s) of metallic hydrogen are (completely) atomic. Such were found in searches with a few or less atoms. The most stable (and nearly degenerate in energy) ones are shown in Fig.\ \ref{fig:atomic_structures}.
%
%
Notice that they are all very similar in their symmetries.

The tendency towards the formation of molecules though can be seen, even in these structures. Notice the close proximity of atoms (into the page); this is approximately $0.9914$ {\AA}. This is analogous to the results for the $c/a \ll 1$ BCT structure (discussed above). This means that these structures too may not be at stable stationary points; confirmed, in results presented below. Further results and discussion about these structures themselves is provided in Section I\! B\! $3$ of the SI.

\subsection{Dynamic Stabilities}

A crystal is dynamically (mechanically) stable, if it executes a stable vibrational motion about its equilibrium configuration \cite{born1954dynamical}. By solving the equations of motion for the normal modes of vibration, this is equivalent to the condition that their frequencies $\omega$ satisfy:
\begin{equation*}
    \omega_j(\mathbf{q})^2 \ge 0
\end{equation*}
for \textit{all} phonon branches $j$ and wavevectors $\mathbf{q}$. (In other words, they are real; for an imaginary frequency means that the system, subject to a small displacement, will disrupt exponentially with time.) Dynamic stabilities were determined by calculating the phonon density of states $F(\omega)$ (along both the real and imaginary axes), for each structure. 

\begin{figure}
    \centering
    \subfloat[stable]{\includegraphics[width=0.45\textwidth]{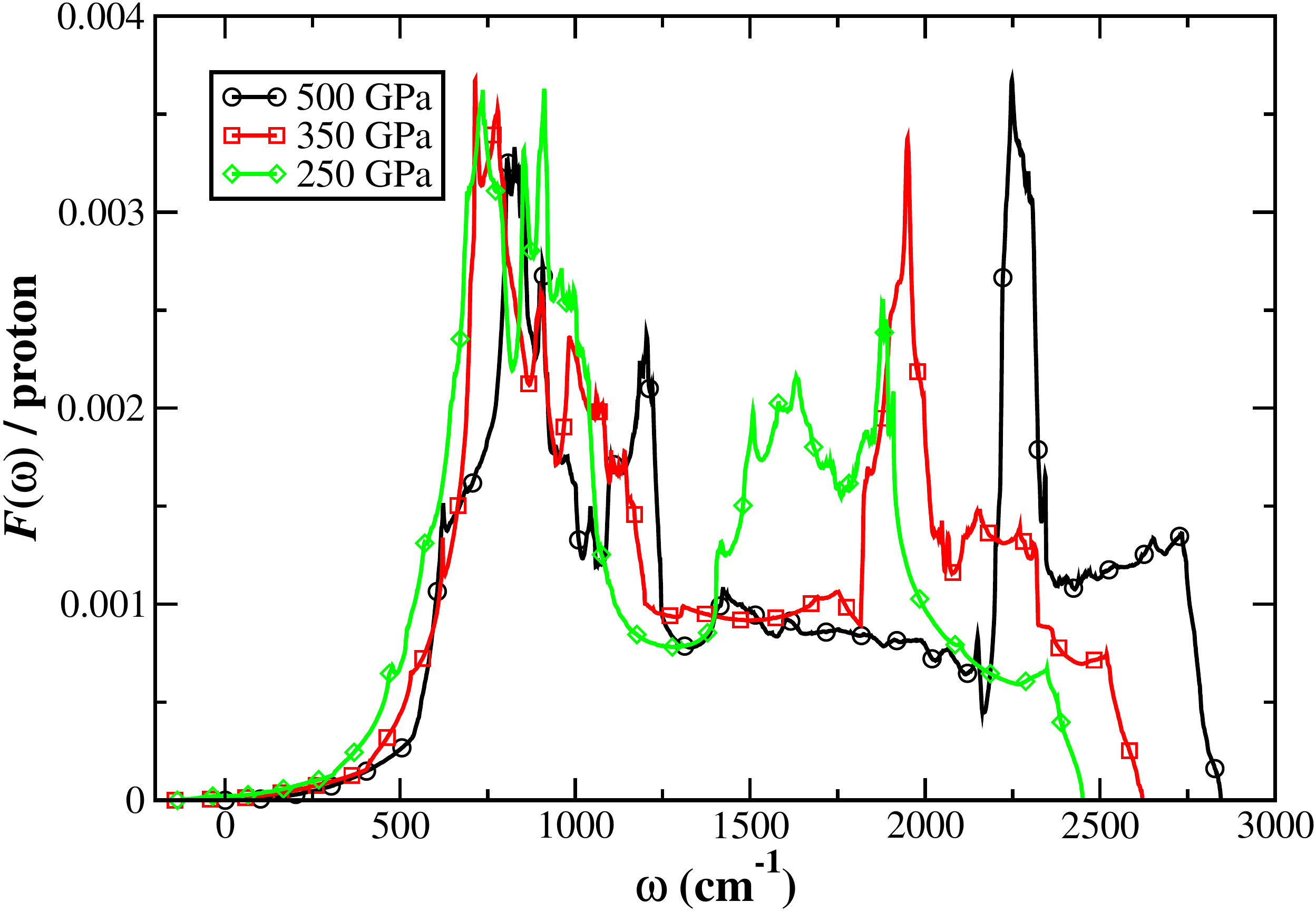}\label{fig:PHDOS_tetragonal:stable}}
    \qquad
    \subfloat[unstable]{\includegraphics[width=0.45\textwidth]{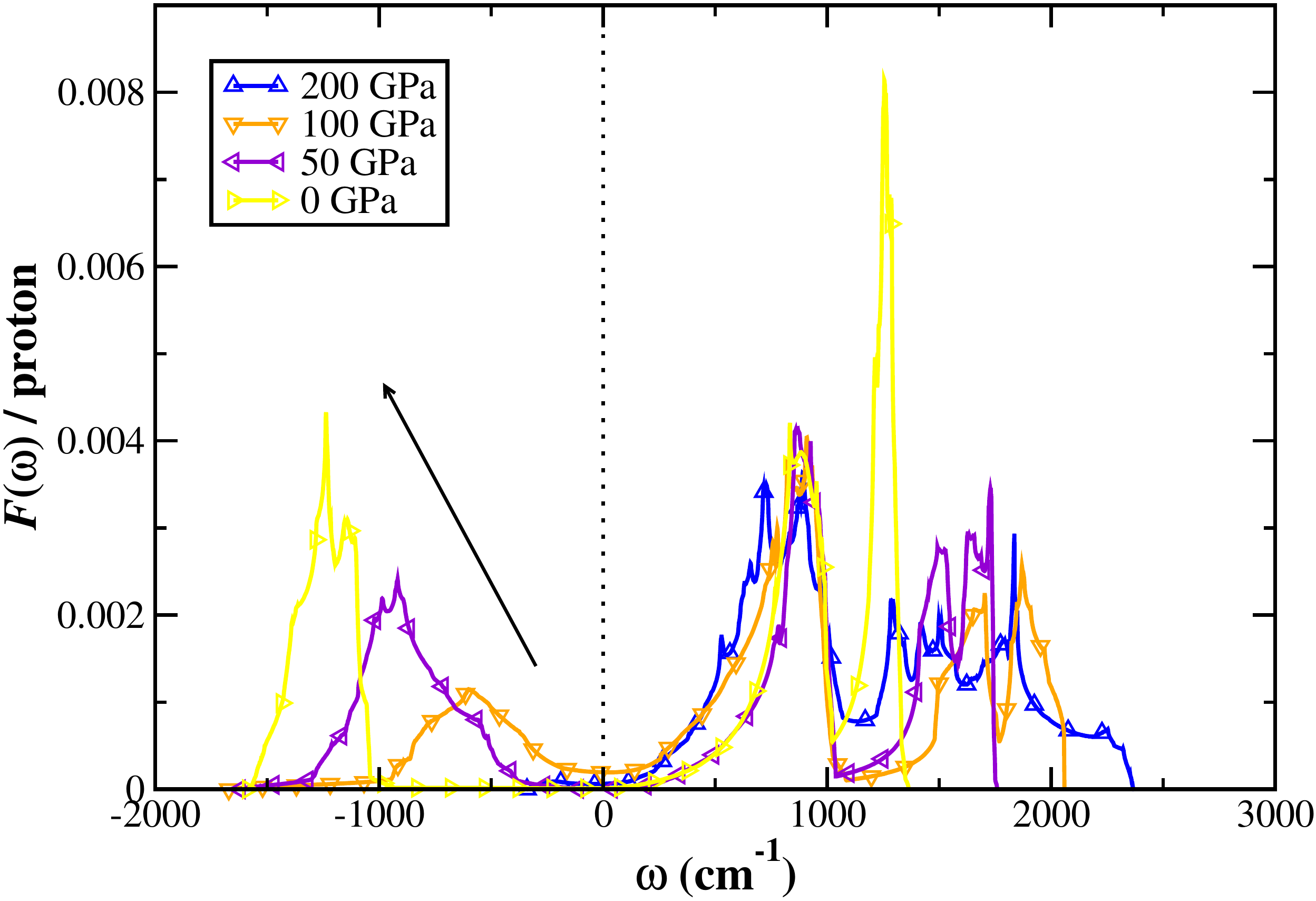}\label{fig:PHDOS_tetragonal:unstable}}
    \caption{Phonon density of states $F(\omega)$ of the BCT Cs-IV structure. Several pressures are shown; separated by regions of (a) stability and (b) instability. The dotted line in (b) is used to separate the stable from unstable (imaginary) frequencies (shown as negative values). The arrow also in (b) highlights the increasing instability, with decreasing pressure.}
    \label{fig:PHDOS_tetragonal}
\end{figure}

Consider first the BCT structures; Fig.\ \ref{fig:PHDOS_tetragonal} shows $F(\omega)$ of Cs-IV.
Note that all other BCT structures are dynamically unstable; see Section I\! C\! $1$ of the SI, for results and discussion about the $\beta$-Sn structure.

Figure \ref{fig:PHDOS_tetragonal}\subref{fig:PHDOS_tetragonal:stable} shows pressures from $500$ (down) to $250$ GPa. The absence of imaginary frequencies proves dynamic stability (at least to these pressures). Consider also the (qualitative) changes in the spectra. To $350$ GPa, they (mostly) retain their shape, with frequencies simply shifting to lower values. Such is expected for a structure that remains qualitatively the same (in terms of its underlying PES). By $250$ GPa, though, while there continues to be a shift to lower frequencies, differences become apparent. 

Figure \ref{fig:PHDOS_tetragonal}\subref{fig:PHDOS_tetragonal:unstable} shows pressures below $250$ GPa. By $200$ GPa, a small (dynamic) instability appears. This then increases, with decreasing pressure. The spectra remain (qualitatively) similar though (at least until $0$ GPa). This behavior, along with that seen in Fig.\ \ref{fig:PHDOS_tetragonal}\subref{fig:PHDOS_tetragonal:stable} (discussed above), is consistent with a transition region between $300$ and $200$ GPa, where the underlying PES of hydrogen changes significantly.

\begin{figure}
    \includegraphics[width=0.5\textwidth]{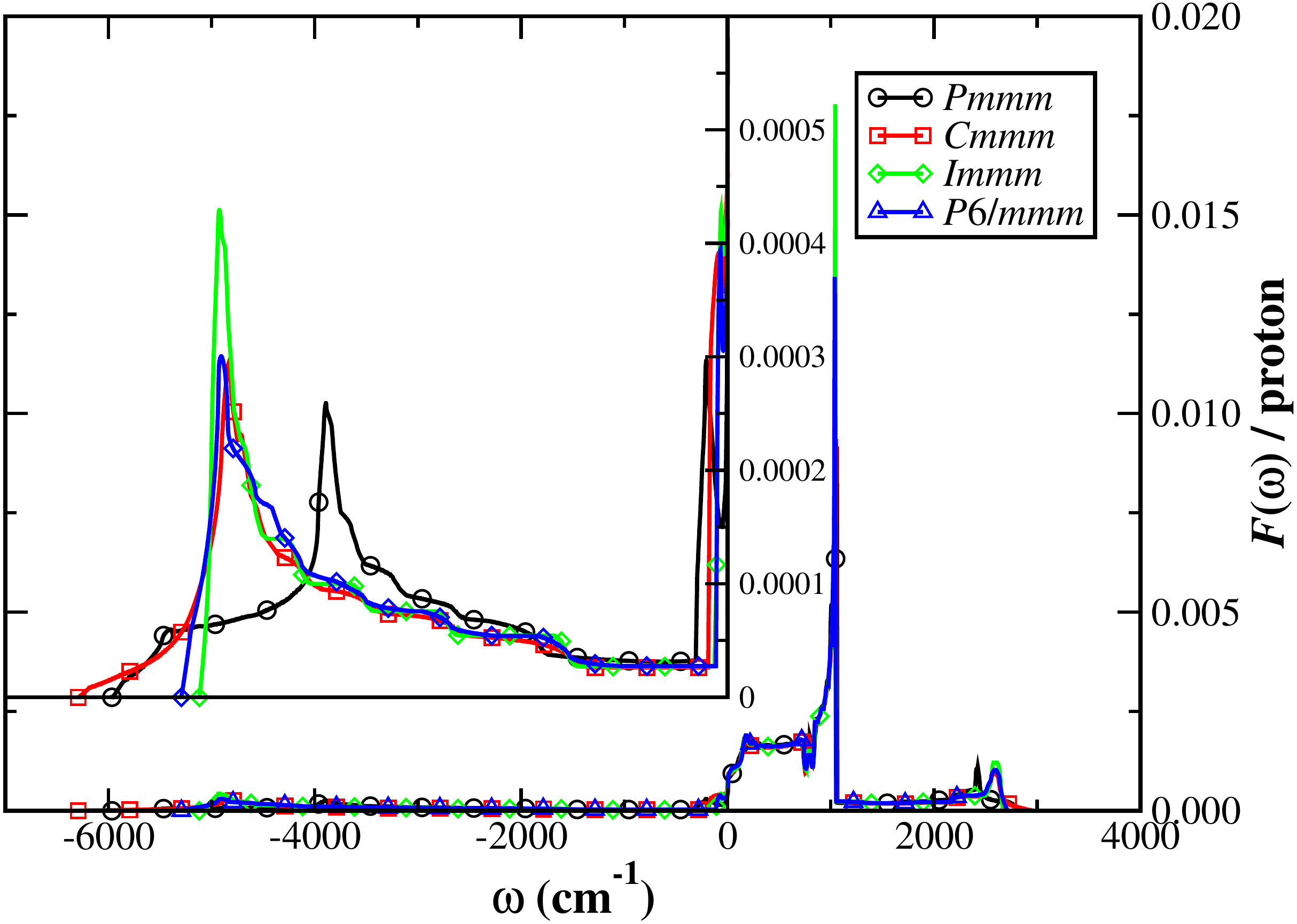}
    \caption{$F(\omega)$ of the candidate (atomic) structures, at $0$ GPa. The inset shows the imaginary-frequency region.}
    \label{fig:PHDOS_1H}
\end{figure}

Consider now the candidate structures; Fig.\ \ref{fig:PHDOS_1H} shows $F(\omega)$ for those shown in Fig.\ \ref{fig:atomic_structures}.
The structures exhibit very similar spectra (as expected, given their similarities). Notice their tails that extend to very high imaginary frequencies. While these have about an order of magnitude smaller magnitude, an expanded view reveals peaks. Therefore, while the most stable candidate (atomic) structures (Fig.\ \ref{fig:atomic_structures}) are at stationary points, they are unstable ones. This is further supported by additional results; see Section I\! C\! $2$ of the SI. Note that also presented and discussed therein are dynamic stabilities of these structures at higher pressures.

\section{Discussion}
\label{sec:discussion}

The possibility of metastable metallic hydrogen has been investigated in detail. This was considered from two approaches, by an analysis of: the BCT structures (stable above molecular dissociation); and the prediction of new candidate ones at $0$ GPa. The results obtained are self-consistent, and provide strong evidence regarding this possibility.

The results in Section \ref{sec:results} are concisely summarized by the processes undergone on the PES of hydrogen, as the pressure is removed (following molecular dissociation):
\begin{enumerate}[{(a)}]
    \item At (relatively) high pressures ($> 300$ GPa): the PES of (atomic) metallic hydrogen is well defined; and it remains qualitatively similar below molecular dissociation.
    \item At intermediate pressures ($200$--$300$ GPa): the energy barriers of metallic hydrogen on the PES collapse; while those of the (completely) molecular phase harden.
    \item At lower pressures ($< 200$ GPa): the PES of molecular hydrogen is well defined.
\end{enumerate}

From these results, we conclude that (atomic) metallic hydrogen is metastable; but only to $300$--$200$ GPa. Below $200$ GPa, metallic hydrogen has \textit{no} region of stability.

Metastable metallic hydrogen, even at these pressures, may have important practical applications. Such pressures are significantly less than those expected for molecular dissociation \cite{PhysRevLett.114.105305}, and are possible to study under static conditions \cite{eremets1996high}. The importance though depends on its properties.

There are therefore several important, open questions. One mentioned, are the properties of metastable metallic hydrogen; for example, superconductivity. Whether this phase is even realizable though depends on its lifetime; and such properties on its finite-temperature stability. Answering these, and related questions, will be the subjects of future work.



\section{Methods}
\label{sec:methods}

\subsection{Electronic-Structure Calculations}

All calculations were performed using the {\qe} ({\qeabbrev}) density-functional theory code \cite{0953-8984-21-39-395502}. The pseudopotential method based on the projector augmented-wave method \cite{PhysRevB.59.1758} was used \footnote{We used the pseudopotential H.pbe-kjpaw\_psl.0.1.UPF from \href{http://www.quantum-espresso.org/}{http://www.quantum-espresso.org/}.}, to replace the bare Coulomb potential of the proton. The Perdew--Burke--Ernzerhof generalized gradient approximation exchange and correlation functional \cite{PhysRevLett.77.3865} was used. {\qeabbrev} is based on a plane-wave basis set, and we used kinetic-energy cutoffs of $57.5$ Ry for the wavefunction and $345.5$ Ry for charge density and potential. For Brillouin-zone sampling, at least $32{\times}32{\times}32$ (unshifted) {\kvec} points were used. The smearing scheme of Methfessel--Paxton \cite{PhysRevB.40.3616} was used for Brillouin-zone integrations, with a smearing width of $0.02$ Ry. These choices give a total convergence, in energy, for example, to better than $0.5$ meV$/$proton.

\subsection{Geometry Optimizations}

Stationary points of structures were found by performing constant-pressure geometry optimizations. These were done using the Broyden--Fletcher--Goldfarb--Shanno algorithm \cite{Fletcher:1987:PMO:39857}, as implemented within {\qeabbrev}. Energies, forces, and pressures were converged to $10^{-5}$ Ry, $10^{-4}$ Ry$/$a.u., and $0.5$ kbar, respectively.

\subsection{Structure Prediction}

Calculations were based on the \textit{ab initio} random structure searching approach \cite{0953-8984-23-5-053201}. These were performed using the electronic-structure calculations and geometry optimizations, as described above. For the former, the basis-set cutoffs were reduced, to $46$ and $221$ Ry, respectively; as was the Brillouin-zone sampling, to $8{\times}8{\times}8$ (shifted) {\kvec} points. For the latter, the energy and force convergence-criteria were relaxed by an order of magnitude. These are (essentially) default values, capable of generating candidate structures for further analysis.

Random structures were constructed by generating random primitive (lattice) vectors, renormalizing the volume (to the expected average), and proton configurations. Searches were performed over unit cells containing $1$ to $10$ atoms. In particular, geometry optimizations (at $0$ GPa) were performed; for each unit-cell size, $1000$ trials. Symmetries of the (relaxed) structures were then determined with the FINDSYM program \cite{Stokes:zm5027}.

\subsection{Dynamic Stabilities}

Phonons were calculated using density-functional perturbation theory, as implemented within {\qeabbrev}. (Phonon) density of states $F(\omega)$ were calculated over a grid of $4{\times}4{\times}4$ phonon wavevectors {\qvec}. Such a {\qvec}-point grid is estimated to be twice as dense (in each direction) as sufficient to accurately calculate $F(\omega)$ of metallic hydrogen \cite{PhysRevLett.106.165302, PhysRevB.84.144515, PhysRevB.85.219902}.




%



\begin{acknowledgments}
J.\ M.\ M.\ acknowledges startup support from Washington State University and the Department of Physics and Astronomy thereat.
\end{acknowledgments}


\

\noindent
\textbf{Author Contributions}

J.\ M.\ M.\ proposed the research. All authors performed the calculations. C.\ M.\ T.\ and J.\ M.\ M.\ analyzed the results. J.\ M.\ M.\ wrote the paper.


\

\noindent
\textbf{Competing Financial Interests}

The authors declare no competing financial interests.




\end{document}